\documentclass[twoside,twocolumn,11pt]{article}
\usepackage{extsizes}
\usepackage[super,sort&compress,comma]{natbib} 
\usepackage[version=3]{mhchem}
\usepackage[left=1.5cm, right=1.5cm, top=1.785cm, bottom=2.0cm]{geometry}
\usepackage{balance}
\usepackage{widetext}
\usepackage{times,mathptmx}
\usepackage{sectsty}
\usepackage{graphicx} 
\usepackage{lastpage}
\usepackage[format=plain,justification=raggedright,singlelinecheck=false,font={stretch=1.125,small,sf},labelfont=bf,labelsep=space]{caption}
\usepackage{float}
\usepackage{fancyhdr}
\usepackage{fnpos}
\usepackage[english]{babel}
\usepackage{array}
\usepackage{droidsans}
\usepackage{charter}
\usepackage[T1]{fontenc}
\usepackage[usenames,dvipsnames]{xcolor}
\usepackage{setspace}
\usepackage[compact]{titlesec}

\usepackage{epstopdf}

\definecolor{cream}{RGB}{222,217,201}

\begin{document}



\makeFNbottom
\makeatletter
\renewcommand\LARGE{\@setfontsize\LARGE{15pt}{17}}
\renewcommand\Large{\@setfontsize\Large{12pt}{14}}
\renewcommand\large{\@setfontsize\large{10pt}{12}}
\renewcommand\footnotesize{\@setfontsize\footnotesize{7pt}{10}}
\makeatother

\renewcommand{\thefootnote}{\fnsymbol{footnote}}
\renewcommand\footnoterule{\vspace*{1pt}%
\color{cream}\hrule width 3.5in height 0.4pt \color{black}\vspace*{5pt}} 
\setcounter{secnumdepth}{5}

\makeatletter 
\renewcommand\@biblabel[1]{#1}            
\renewcommand\@makefntext[1]%
{\noindent\makebox[0pt][r]{\@thefnmark\,}#1}
\makeatother 
\renewcommand{\figurename}{\small{Fig.}~}
\sectionfont{\sffamily\Large}
\subsectionfont{\normalsize}
\subsubsectionfont{\bf}
\setstretch{1.125} 
\setlength{\skip\footins}{0.8cm}
\setlength{\footnotesep}{0.25cm}
\setlength{\jot}{10pt}
\titlespacing*{\section}{0pt}{4pt}{4pt}
\titlespacing*{\subsection}{0pt}{15pt}{1pt}


\makeatletter 
\newlength{\figrulesep} 
\setlength{\figrulesep}{0.5\textfloatsep} 

\newcommand{\topfigrule}{\vspace*{-1pt}%
\noindent{\color{cream}\rule[-\figrulesep]{\columnwidth}{1.5pt}} }

\newcommand{\botfigrule}{\vspace*{-2pt}%
\noindent{\color{cream}\rule[\figrulesep]{\columnwidth}{1.5pt}} }

\newcommand{\dblfigrule}{\vspace*{-1pt}%
\noindent{\color{cream}\rule[-\figrulesep]{\textwidth}{1.5pt}} }

\makeatother

\twocolumn[
  \begin{@twocolumnfalse}

 \noindent\LARGE{\textbf{Homogeneous Alignment of Liquid Crystalline Dendrimers Confined in a Slit-Pore: Computational Simulation Study.}} \\

 \noindent\large{Zerihun G. Workineh$^{\ast}$ and Alexandros G. Vanakaras} \\

\noindent\normalsize{In this work we present results from $NPT$ (isobaric-isothermal) Monte Carlo Simulation studies of Liquid Crystalline Dendrimer (LCDr) systems confined in a slit-pore made of two parallel flat walls. We investigate the substrate induced conformational and alignment  properties of the system at different thermodynamic state points under uniform (unidirectional) anchoring condition. Tractable coarse grained force fields to model both monomer-monomer and monomer-substrate interaction potentials have been used from our previous work. In this anchoring condition, at lower pressure almost all the monomers are anchored to the substrates and mesogens are perfectly aligned with the aligning direction. This alignment is not uniformly transmitted to the bulk region as the pressure grows, instead, it decays with distance from the surface to the bulk region. Due to this reason, the global orintational order parameter decreases with increasing pressure (density). In the neighborhood ($2-3$ mesogenic diameter) of upper and lower walls, mesogenic units form smectic A like structure whose layers are separated by layers of spherical beads. In this region individual LCDrs possess a rod like shape.} \\


 \end{@twocolumnfalse} \vspace{0.6cm}

 ]

\renewcommand*\rmdefault{bch}\normalfont\upshape
\rmfamily
\section*{}
\vspace{-1cm}


\footnotetext{\textit{Department of Materials Science, University of Patras, Patras, Greece. Fax: +30-2610-996156; Tel: +30-2610-996156; E-mail: workzer@upatras.gr}}




\section{\label{sec:level1}Introduction}
 Liquid Crystalline Dendrimers (LCDrs) are part of dendrimers which usually are derived through functionalization of common dendrimers with low molar mass liquid crystal molecules\cite{lc16, lc17,lc18,Goodby1998}. Self-assembly and phase behavior of the combined supermolecules significantly depend on the type of LC-dendrimer connection\cite{lc18,lc19,lcp1,lcp2,lcp3}. Inclusion of anisotropic molecules in to nearly spherical dendritic scafolds enables to obtain multifunctional anisotropic macromolecules. This opens up the possibility of achieving a wide variety of physical properties with tailored made potential applications. Consequently, rigorous research has been conducted in the last few years on the synthesis and characterisation \cite{lc16,lc17,lc18,lc19, Kim2013} as well as in theory\cite{terzis_conformational_2000, vanakaras_ordered_2001,lc253}, modeling and computational simulation\cite{lc33,lc34,vanakaras_2005,lc251,lc252} of these materials. 
 
 The interplay between the tendency of maintaining spherically symmetric conformation due to flexible core segments and tending towards anisotropic conformation as the result of peripheral mesogenic units determines the overall structural and dynamic properties of bulk LCDr systems. Strong competition between these two factors could be influenced by the kind of pair interaction potential energies of dendritic segments and thermodynamic conditions on which the system is imposed. The phase behavior of bulk or confined LCDr system could also be highly affected by these two factors\cite{lcp1,lcp2,lcp3}. Mark Wilson and coworkers have proved that bulk LCDr system exhibits isotropic, nematic and smectic phases depending on the length of mesogenic units and the temperature\cite{lc33,lc34}. According to their findings, as the system goes from isotropic state to anisotropic state, single LCDr molecules could be changed from spherical shape to either rod-like or disc like shapes depending on the kind of dendrimer-LC connectivity (side-on or end-on).

The study of ordered stable structures for a system of particles interacting through a specific kind of interactions is an interesting problem both from the scientific and technological point of views. This problem is still active research curiosity. The search for its comprehensive and satisfactory solution poses a formidable challenge of academical and technological interests. Looking for these ordered equilibrium structures becomes more intricate when their system is under a certain geometric constraints due to an external confinements\cite{mc2007}. 

The problem becomes more complicated when it comes to liquid crystalline dendrimer systems under external factors imposed by the mean field or by a confining geometry. Confinement of molecular phases in a porous medium usually modifies their overall properties, in terms of structure, phase
behavior and molecular dynamics \cite{lc1,lc2}. The sole reason behind these all changes of properties arises from delicate energetic balance between adhesive(LCDr-surface interaction) and cohesive(inter/intra dendritic interaction) forces, in addition to thermodynamic conditions. The global and local structural and alignment properties of confined LCDr system can be enormously affected by the strength of surface anchoring potential. Confining surface with weaker adhesive force could affect positions and orientations of only few molecules which are in the first vicinity. On the other hand, those with strong adhesive force can induce long ranged effect on the system.  

 Liquid Crystals (LCs) are elastically soft materials with long range translational and orientational orders that may directly couple to the surface of the porous medium \cite{lc3,md2014}. The various possible arrangements adopted by LC systems of micro or macro scales with different shapes of confining pores and surface interactions have been widely investigated experimentally and quite well understood in the frame of continuum elastic theories, including the occurrence of director distortions or topological defects \cite{lc4,tr1992}. Through controlling the surface-LC interactions, usually by means of chemical and/or mechanical treatment of the substrate, a variety of alignments (homeotropic, planar, tilted, etc) of the LC medium with respect to the substrate are possible. However, in the case of LCDrs, the mechanism behind surface anchoring and alignment do not involve only the translational and orientational restrictions imposed by the substrate to its vicinal beads/mesogens but also the positional/orientational correlations among the dendritic segments. The stability of ordered mesophases could be enhanced or reduced by the presence of confining pores and it strongly affects the nature of the phase transitions \cite{lc5,lc6,lc7,lc8,lc9}. Computer simulation methods are quite unique tools to investigate the structure of nanoconfined LC molecules \cite{lc10,md2011,md2013,md2009,mc2005,mc2008,mc2011}. They provide direct insight into the interfacial properties and the inhomogeneous nature of the confined phases. 

Managing the macroscopic alignment of these anisotropic materials in a confinement is a key factor for many of their potential applications. Thus, in a system of confined liquid crystals, the first aligned layer in the vicinity of the substrate would be used as the template to induce the next layer so that alignment could propagate through out the range of the confined film depending on the temperature and the density of the system. However, in LCDrs, that seems unlikely due to the fact that the first layer of mesogenic units in the neighbourhood may be shielded by the core spherical beads under bead-wall repulsive potential circumstances. In this case, for the reason of chemical incompatibility, mesogens will avoid to get closer to the layer of beads above the first layer of mesogens. This breaks alignment correlations between the first and the next layers of mesogens. However, this might be resolved by making the confining wall attractive to both mesogens and beads. In this case, bead layer and mesogen layer would be formed in the first neighbourhood, not one over the other but in parallel. Thus, these first layers of each type could be templates to induce further layers. This situation is different for different anchoring conditions. The extent of layer propagation from the wall surface to the bulk region depends not only on the kind of anchoring conditions, it also depends on anchoring strength, monomer-monomer interaction potential and the thermodynamic conditions. 

Several different models for common dendrimers (isolated and confined)\cite{lc28, lc29, lc30,lc31,lc32}, liquid crystal dendrimers\cite{lc33,lc34, lc35,lc36,lc37} and dendronized polymers\cite{christopoulos_structure_2003, christopoulos_helix_2006, Cordova-Mateo2014} have been proposed and used for computer simulation studies of their properties. These models range from detailed atomistic to coarse grained ones. In atomistic models, detailed interaction potentials between individual atoms (molecules) should be considered rendering them as computationally expensive models. Alternatively, in coarse grained models groups of atoms are represented as united interacting sites, preserving at the same time the architectural characteristics of the dendrimer. 

The main purpose of this work is to demonstrate Monte Carlo simulation results of model LCDr system confined under a slit pore made of two parallel walls whose inner surfaces induce homogeneous planar alignment to the mesogens based on our previous work\cite{zerihun}. Mainly, structural and alignment properties of the target system have been reported. As far as our knowledge, no study results of LCDrs under confinement have been reported so far.  The rest of the paper is organized as: In section II models and details of our simulation are presented and simulation results and their discussions are revealed in section III. Finally, we wrap up our work with conclusions and remarks in section IV.  

\section{\label{sec:level1} Simulation model and details}
\begin{figure}[h!]
\centering
\includegraphics[width=3.0in]{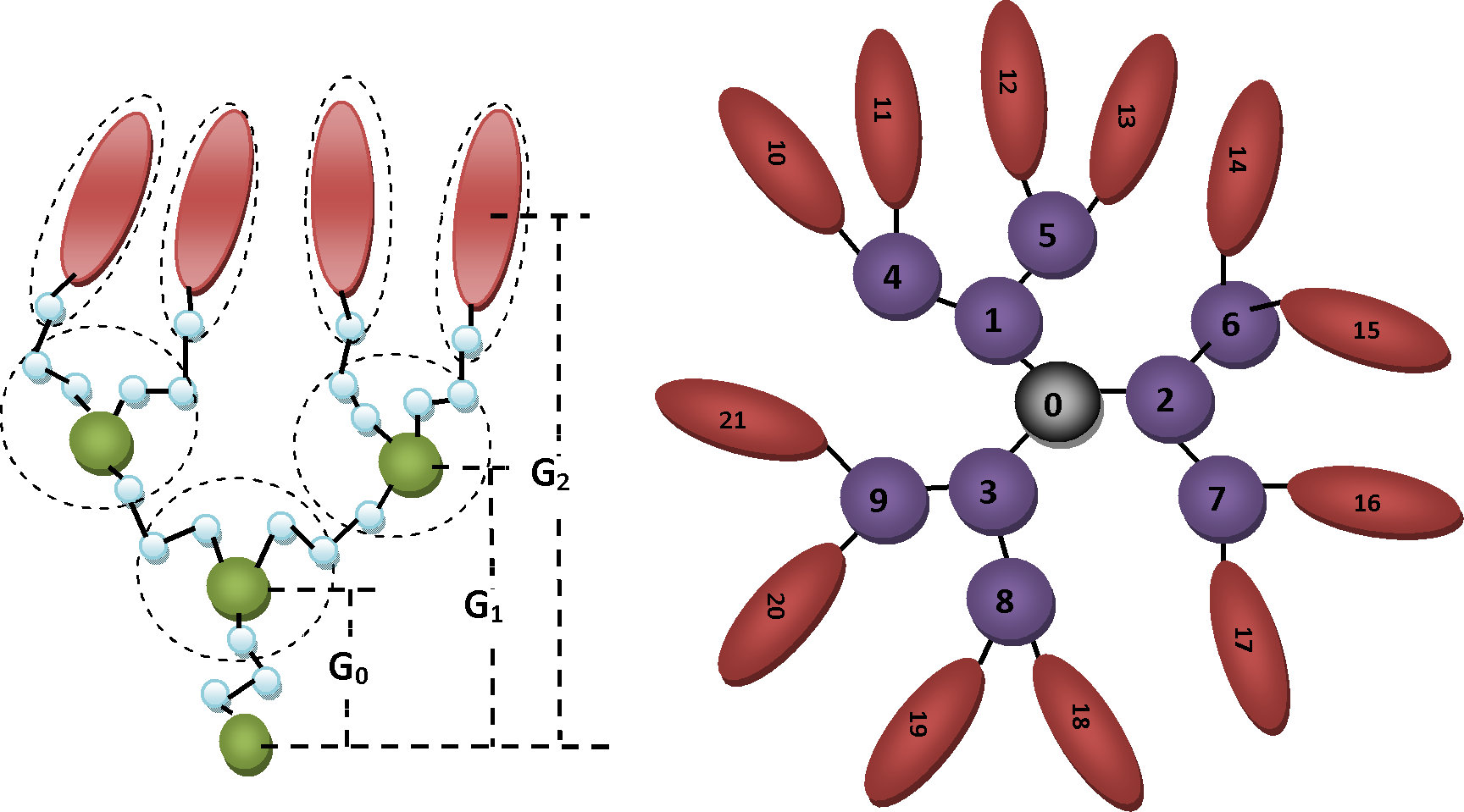}
\caption{Model $2^{nd}$ generarion Liquid Crystal dendron with $f_c=2$ and spacers (left) and a coarse-grained dendritic supermesogen composed of 3 dendrons (right) ($G_{2}D_{3}$) } 
\label{fig:fig1}
\end{figure}

In this study, we consider a system of generation \emph{two}, core functionality \emph{three} and branch functionality \emph{two} model LCDr (denoted as $G_2D_3$) confined in a slab geometry. 
The model single LCDr, its intra dendritic interaction potential energy and LCDr-substrate interaction potential energy have been reported in our previous work\cite{zerihun}. Our current study is mainly based on the models developed in this reference. According to this model, the junction units are united atoms which represent collectively the atoms around each branching points. The junctions are connected with virtual bonds with variable length which preserve the architectural connectivity of the dendrimer allowing at the same time the junction units to sample available free volume in the interior of the dendrimer. With this assumption the dendrimer is composed of two different spherically symmetric sites (denoted with $b$) representing the junction points and one representing the core of the dendrimer. The mesogenic units (denoted with $m$) are assumed to be cylindrically symmetric and are connected by one of their ends to the $G-1$ generation junction beads with virtual bonds (see Fig.~\ref{fig:fig1}(left)). As a result, a dendritic conformation is fully described with the positions of the junction sites $\mathbf{r}_b$ and the positions and orientations of the mesogenic units $\mathbf{r}_m$ and $\mathbf{\hat{u}}$, respectively. 

To model homogeneous anchoring conditions of the LCDr system confined in a slit pore made of two parallel impenetrable walls, $N=108$ equilibrated $G_2D_3$ LCDrs have been duplicated throughout the volume of the slit according to cubic lattice arrangement. Duplication is made in such a way that the distance between any two LCDrs is far enough to consider the inter-LCDr potential energy to be $zero$, initially. The two confining parallel walls are placed at $z=0$ (bottom wall) and $z=L_z$ (top wall). Initially, any LCDr is placed well above the bottom wall and below the top wall in order to avoid initial segment-surface overlap. 

In the level of structural resolution described above, the total intra/inter-molecular potential energy of LCDr system is the sum of the bonded ($B$) and non-bonded ($N\text{-}B$) pair interactions, 
\begin{eqnarray}
U& = &U^{B} + U^{N\text{-}B}\nonumber\\
 & = &\sum_{i,j}U^{B}(l_{ij})
 + \sum_{i,j}U_{pq}^{N\text{-}B}(\mathbf{\hat{r}}_{ij},\mathbf{u}_{i},\mathbf{u}_{j}) 
\label{eq:eq1}
\end{eqnarray}
\noindent
here $l_{ij}$ is the length of the virtual bond connecting the junctions $i,j$. The indexes $p,q$ in $U_{pq}^{N\text{-}B}$ may be either $b$ or $m$ denoting junction beads and mesogenic segments, respectively. $\mathbf{r}_{ij}$ is the vector that connects the centres of the non-boded sites $i,j$ and the unit vector $\mathbf{u}_{i}$ denotes the orientation of the $i^{th}$ mesogenic unit.
    
For the bonded potential we have adopted a simple pair potential which corresponds to a freely fluctuating bond;
 \begin{equation}
 U^{B}(l) = \left\{\begin{array}{rl}
0,  &  l_{min}<l<l_{max}\\ 
\infty,  & \text{otherwise},
\end{array} \right.
\label{eq:eq2}
\end{equation} 
\noindent
where $l_{min}$ and $l_{max}$ are the minimum and maximum allowed separation distances between two bonded segments. The interaction potential of two non-bonded sites of type $p$ and $q$ is given by

\begin{equation}
U_{pq}^{N\text{-}B}(\textbf{r}_{ij},\textbf{u}_{i},\textbf{u}_{j}) = \left\{\begin{array}{rl}
U^{N\text{-}B}_{bb}(r_{ij}),  & p=q=b\\ 
U^{N\text{-}B}_{mm}(\hat{\mathbf{r}}_{ij},
\textbf{u}_{i},\bf{u}_{j}), &  p=q=m\\ 
U^{N\text{-}B}_{bm}(\mathbf{\hat{r}}_{ij},\textbf{u}_{j}), &  p=b,q=m
\end{array} \right.
\label{eq:eq3}
\end{equation} 
\noindent
The spherical junction segments are modeled as Lennard-Jones spheres interacting through ($b-b$ interaction) as,
\begin{equation}
U^{N\text{-}B}_{bb}(r_{ij}) = 4\epsilon_{0bb}\left[\left(\frac{\sigma_{0bb}}{r_{ij}}\right)^{12} -\left(\frac{\sigma_{0bb}}{r_{ij}}\right)^{6}\right],
\label{eq:eq4}
\end{equation}
\noindent
and the mesogens are modelled as cylindrically symmetric soft ellipsoids interacting with the widely used Gay-Berne interaction potential\cite{Gayberne},
\begin{equation}
U^{N \text{-} B}_{mm}  = 4\epsilon_{mm}(\mathbf{\hat{r}}_{ij},\mathbf{u}_{i},\mathbf{u}_{j})\left[\left(\frac{\sigma_{0mm}}{r}\right)^{12}
- \left(\frac{\sigma_{0mm}}{r}\right)^{6}\right],
\label{eqn:5}
\end{equation}
\noindent
with $r=(r_{ij}-\sigma_{mm}(\hat{\mathbf{r}}_{ij},\mathbf{u}_{i},\mathbf{u}_{j}) + \sigma_{0mm})$. Mathematical descriptions of relative orientation dependent quantities, $\sigma_{mm}$ and $\epsilon_{mm}$ and some associated parameters can be refereed from our previous work\cite{zerihun}.
\noindent
\noindent
\noindent
\noindent


 The bead-mesogen interaction potential is modelled by the GB potential of equation \ref{eqn:5}, where either $\textbf{u}_{i}$ or $\textbf{u}_{j}$ is considered zero and based on the parametrization shown in the Table $1$.

Interactions of a Gay-Berne particle with solid surfaces have been described in several ways \cite{zerihun, lc46, lc47,lc48,lc49}. In this study we use the interaction model introduced in reference\cite{zerihun} and references therein. According to this model, each dendritic segments (beads or mesogenic units) of an LCDr with coordinates $(x,y,z)$ interacts with one of the two phantom mesogens centred at either $(x,y,L_z)$ in the top wall or $(x,y,0)$ in the bottom wall. The orientation of the phantom mesogens with respect to the surface determines the anchoring conditions the substrate imposes to the LCDrs. In this study only uniform (unidirectional) planar anchoring condition is considered therein assuming that the phantom mesogens point along a given direction on the surface which without loss of generality is chosen to be the x-axis.

The mesogen-surface interaction effective potential is given by anisotropic $9-3$ Lennard-Jones potential;
\begin{equation}
U^{mw}  = \frac{2\pi}{3}\epsilon_{mw}(\mathbf{\hat{r}}_{ii'},\mathbf{u}_{i},\mathbf{u}_{i'})
\left[\frac{2}{15}\left(\frac{\sigma_{0mw}}{r_{eff}}\right)^{9}
 - \left(\frac{\sigma_{0mw}}{r_{eff}}\right)^{3}\right],
\label{eq:eq6}
\end{equation}
where, $r_{eff}=r_{ii'}-\sigma_{mw}(\mathbf{\hat{r}}_{ii'},\mathbf{u}_{i},\mathbf{u}_{i'}) + \sigma_{0mw}$, $\mathbf{u}_{i}$ and $\mathbf{u}_{i'}$ are unit vectors along the long axes of mesogenic unit $i$ and phantom mesogen $i'$. $\mathbf{r}_{ii'}$ is the intermolecular vector which connects the center of mass of a mesogenic unit with its phantom counterpart which is located at less distant point of the surface to the actual mesogen. For unidirectional planar anchoring conditions, $\mathbf{u}_{i'} = \mathbf{\hat{x}}$. Exactly the same equation as Eq. \ref{eq:eq6} is used to model bead-wall interaction potential energy except that $\mathbf{u}_{i}=0$ in this case. Mathematical definitions for $\sigma_{mw}$ and $\epsilon_{mw}$ are similar to $\sigma_{mm}$ and $\epsilon_{mm}$, respectively, whose associated constant parameters are given in Table $1$.  

\begin{figure}[h!]
\centering
\begin{tabular}{cc}
\includegraphics[width=1.5in,height=1.2in]{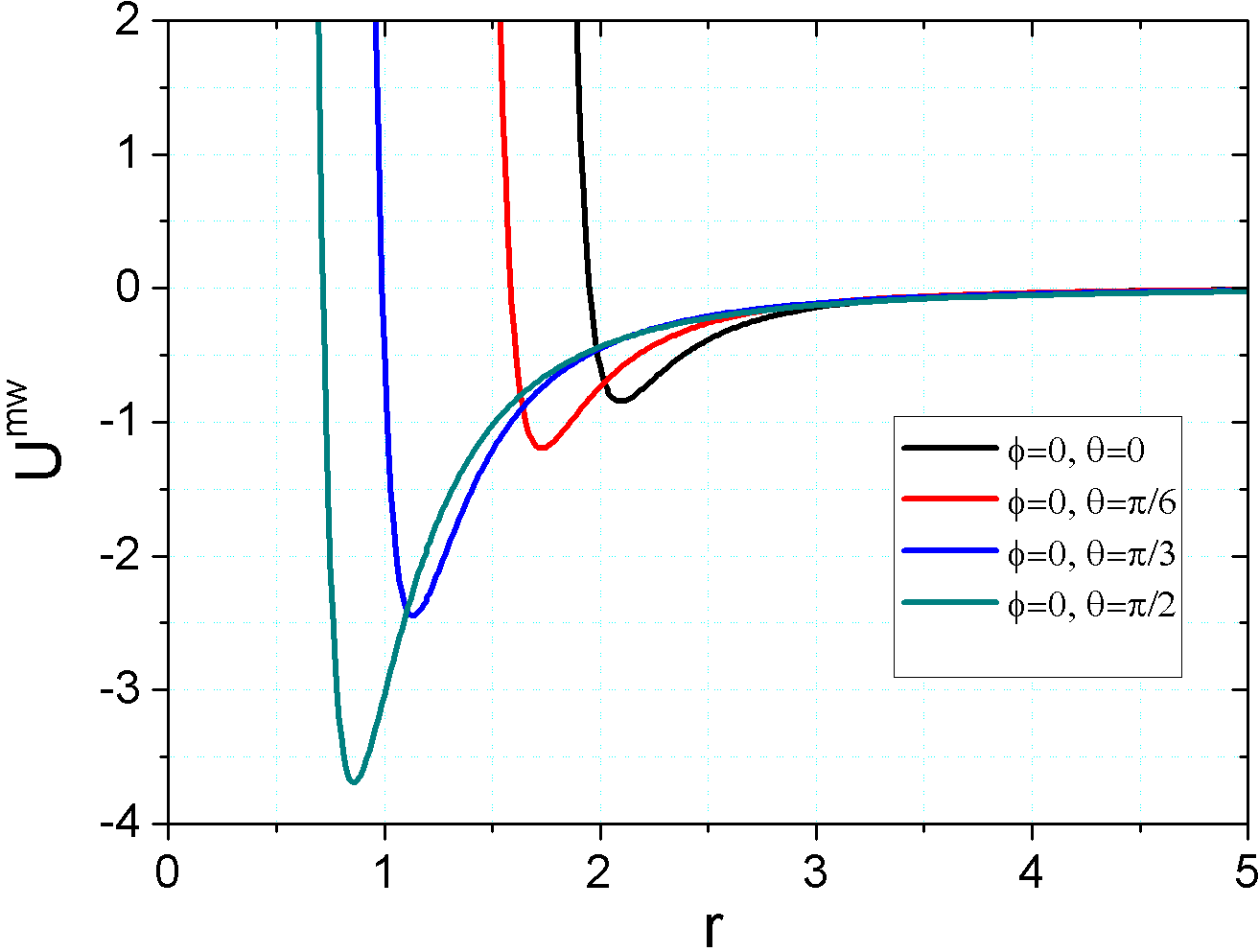}
\includegraphics[width=1.5in,height=1.2in]{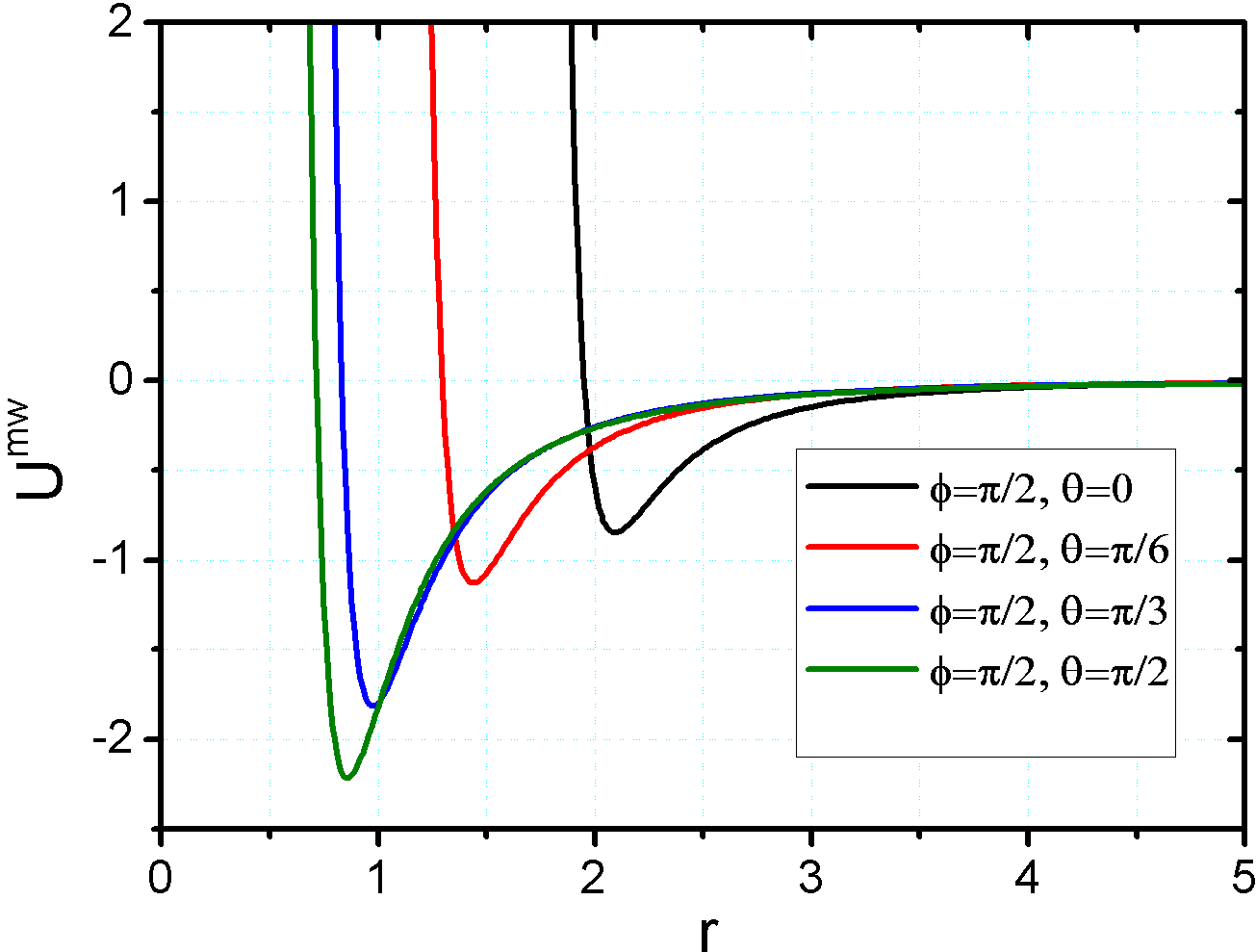}\\
\end{tabular}
\caption{Mesogenic unit-surface potential at four specific values of the polar angles $\theta$ between the mesogenic axis and the surface normal as the function of distance in unidirectional planar anchoring condition at (left) $\phi=0$ and (right) $\phi=\pi/2$. }
\label{fig:fig3}
\end{figure}

Plots of the mesogen-surface interaction potential as a function of the distance, $r$ of the mesogenic unit from the surface for unidirectional planar alignments are shown in Fig.~\ref{fig:fig3} for chosen mesogen-surface configurations at $\phi=0$ and $\frac{\pi}{2}$.   
According to this plot, azimuthally parallel configurations of real mesogens and phantom mesogens are energetically favoured in comparison to perpendicular arrangements at higher polar angles. Nevertheless, at lower polar angles, variation in azimuthal angle hardly affects surface-mesogen interaction potential. Controlling parameters, $\nu_{w}= 1$,  $\mu_{w}= 2$, $\chi_{w} =0.8$ and $\chi^{'}_{w}=0.382$ are used to produce Fig.~\ref{fig:fig3} and are used throughout this work. 
\begin{table}
\centering
\caption{ Force field parameters for the coarse grained LCDr model} 
\centering 
\begin{tabular}{ccccc}
\hline\hline 
 Parameter & Description & values\\ [0.8ex] 
\hline 
D=$\sigma_{0mm}$ & length unit & 1.0 \\
L & length of mesogen & $3D $ \\ 
$\sigma_{0bb}$ & b-diameter& $1.2D$ \\
$\sigma_{0bm}$ & m-b diameter & $\frac{\sigma_{0bb} + \sigma_{0mm}}{2}$ \\
$\sigma_{0bw} $ & b-wall diameter & $\sigma_{0bm}$ \\
$\sigma_{0mw}$ & m-wall diameter & $\sigma_{0mm}$ \\
$\epsilon_{0mm}$ & energy unit &  1\\
$\epsilon_{0bb}$ & b-b interaction & 0.5$\epsilon_{0mm}$  \\
$\epsilon_{0bm}$ & m-b interaction & $\sqrt{\epsilon_{0mm}\epsilon_{0bb}}$  \\
$\epsilon_{0bw}$ & b-wall interaction & $\epsilon_{0bm}$  \\
$\epsilon_{0mw}$ & m-wall interaction & $\epsilon_{0mm}$  \\
$l_{max}$ & max. bond length & $1.8\sigma_{0mm}$ \\
$l_{min}$ & min. bond length & $1.2\sigma_{0mm}$ \\
\hline 
\end{tabular}
\label{table:nonlin} 
\end{table}
 
 Analysis was performed by dividing stored system configurations
into $L_z/\delta L$ equidistant slices ($L_z$ is the slit gap and $\delta L$ is slice width) and calculating averages of relevant observables in each slice. This yielded profiles of quantities such as number density, $\rho^{w}(z)$ ($w$ refers to either bead or mesogen), from which structural changes could be assessed. Number density profile along surface normal (z-axis) direction could be calculated as, 
\begin{equation}
 \rho^{w}(z) = \frac{1}{N^w}\left\langle \sum_{i}{\delta{(z - \mathbf{r}_i\cdot\hat{\mathbf{z}})}}\right\rangle,
 \label{eq:eq11}
\end{equation} 
where, $\mathbf{r}_{i}$ is the position vector of site $i$ and, $N^w$ is the total number of beads ($N^b$) or mesogens ($N^m$). Orientational order profiles were also calculated, particularly from orientational order tensor as;
\begin{equation}
 S_{\alpha \beta}(z) = \frac{\left\langle \mathbf{Q}_{\alpha \beta}(z)\right\rangle}{\left\langle N^m(z) \right\rangle},
 \label{eq:eq12}
\end{equation} 
where, $\mathbf{Q}_{\alpha \beta}(z)=\sum_{i}{(3(u_{i,\alpha}\cdot u_{i,\beta})^2-\delta_{\alpha \beta})}$ and $N^m(z)$ is the instantaneous number of mesogens occupying the relevant slice of volume, $V_z= L^2 \delta L$ with constant slice width of $\delta L$. $S_{\alpha \beta}(z)$ measures orientational order variation across the confined films with respect to the substrate normal.

 Standard Metropolis Monte Carlo (MC) computer simulation has been used to investigate structural behaviour and the possibility of alignment transmission from the surface to the bulk in a system of LCDrs confined in a slit pore. MC simulations have been performed in isobaric-isothermal (NPT) statistical ensemble. The confining box has dimensions $L_x=L_y=L\sigma_{0mm}$ and $L_z=20\sigma_{0mm}$. The periodic boundary conditions along the x-and y-directions give an infinite slit pore. Three systems were initially prepared at reduced pressure, $P^*=P\sigma_{0mm}/\epsilon_{0mm}=0.0$ and reduced temperatures, $T^*=Tk_B/\epsilon_{0mm} =1.0$, $1.5$ and $2.0$. Then, each systems are compressed with pressure increment of $0.05$ up to $P^*=P^*_{max}$ at constant respective temperatures. The volume compression is made along the x- and y-directions keeping the slit gap constant. At each pressure/temperature MC simulations with $10^6$ equilibration and $10^6$ production cycles have been performed. In an MC cycle one random displacement on average is attempted for each molecular segment in addition to one random reorientation for the mesogenic units. Finally, we allow the volume of the box to change ($\Delta V = V_n-V_o$) according to isobaric-isothermal distribution function $f_{NPT}\propto e^{-(U_{tot} + PV)/k_BT}$ with the acceptance probability,
  \begin{equation}
 P_{acc} = Min\left[1, e^{-\beta \left[ PL_z(L_n^{2}-L_o^{2}) + (U_n - U_o) - \frac{2N+1}{\beta}ln\left(\frac{L_n}{L_o}\right)\right]}\right], 
 \end{equation}
 where $L_n$ and $L_o$ are new and old lengths, respectively, of the box along x-and y-directions.
The range of random translations/reorientations and volume change are tuned to give an overall acceptance ratio of about 33\%.

 \section{\label{sec:level1}Simulation Results}  
\begin{figure}[h!]
\centering
\begin{tabular}{cc}
\includegraphics[width=3.0in,height=2.0in]{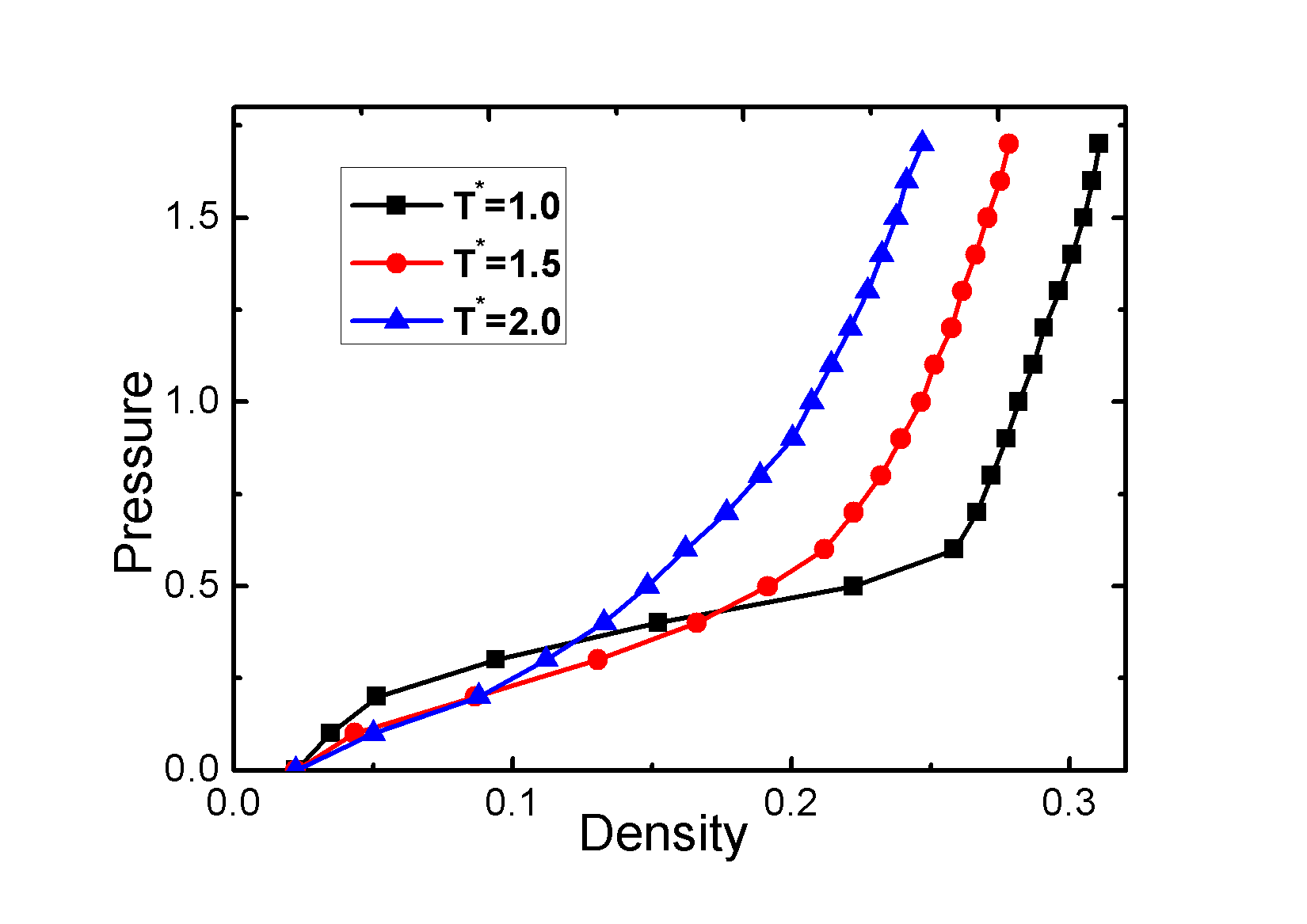}\\
\end{tabular}
\caption{Pressure-density phase diagram for the model LCDr system confined in a slit-pore under uniform planar anchoring conditions at three different temperature values.} 
\label{fig:fig4}
 \end{figure}

\begin{figure}[h!]
\centering
\begin{tabular}{cc}
\includegraphics[width=1.5in,height=1.2in]{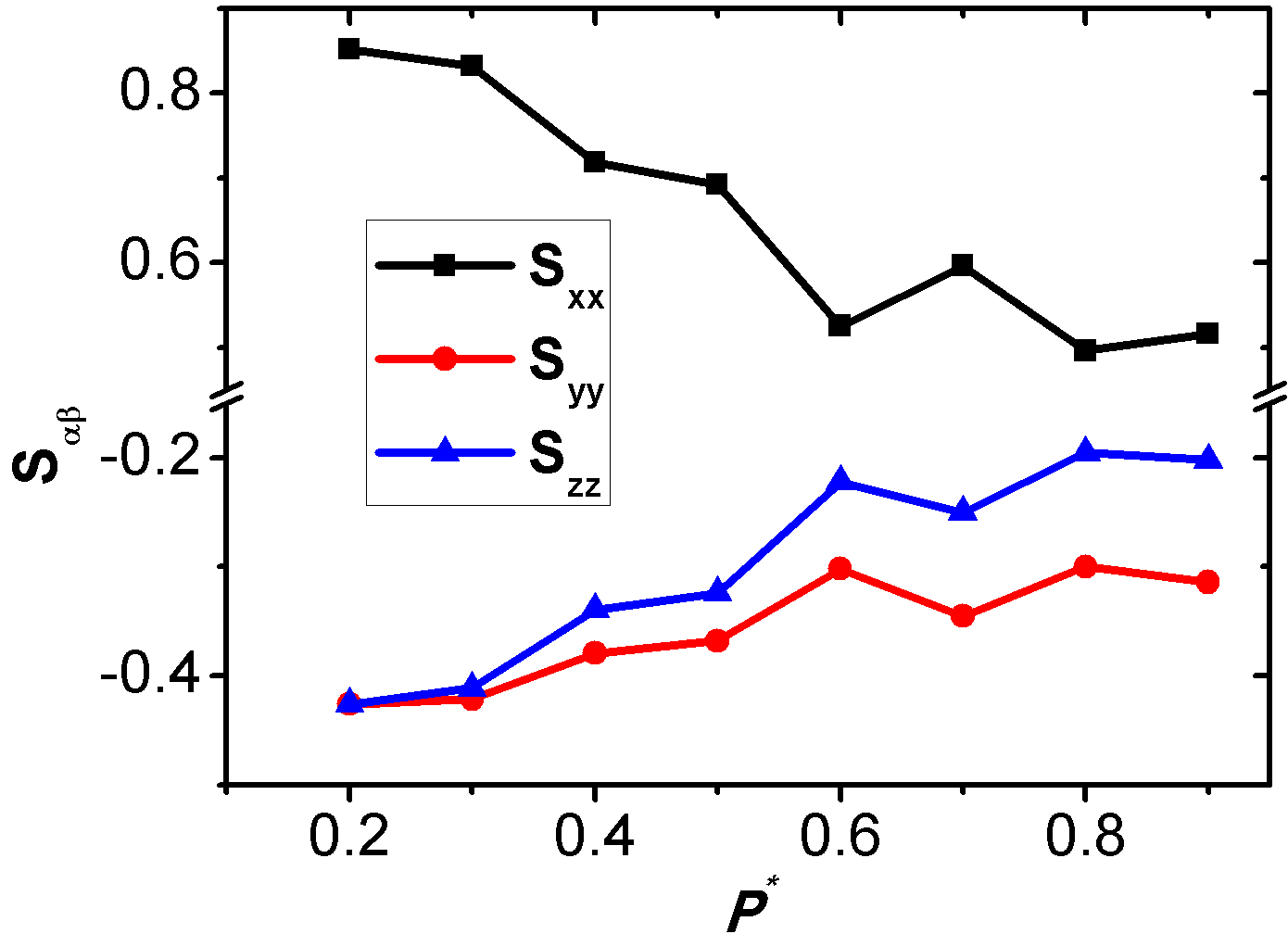}
\includegraphics[width=1.5in,height=1.2in]{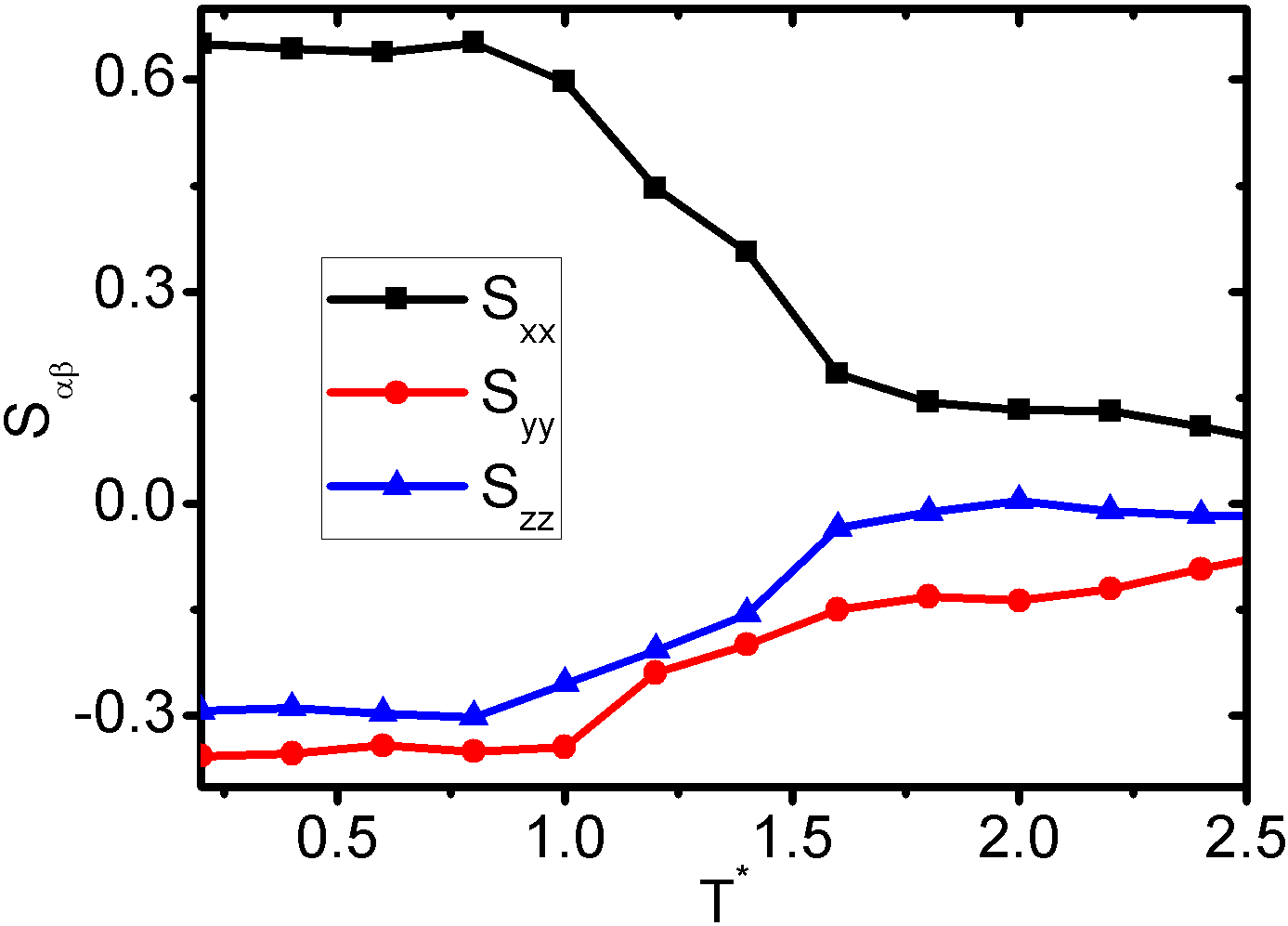}\\
\end{tabular}
\caption{Global values of the components of orientational order tensor as the function of pressure (left, fixed temperature, $T^*=1.0$) and temperature (right, fixed pressure, $P^*=0.6$).} 
\label{fig:fig5}
 \end{figure}

The pressure-density phase diagram of model LCDr system confined in a nano-pore of parallel walls under uniform planar anchoring condition is shown in Fig.~\ref{fig:fig4}. From this figure one can easily notice that there is abrupt change in density of the system at lower temperature unlike its relatively smooth transition at higher temperature. It seems up to a certain value of pressure, the configuration where almost all LCDrs are residing in the neighbourhood of the walls was energetically comfortable at lower temperature. After a given \emph{critical} pressure, density suddenly starts to increase as the result of LCDrs leaving the surfaces of the walls and filling the free space due to compression.    
The global orientational order parameters along each directions, derived from orientational order tensor shown in Eq.\ref{eq:eq3}, are also depicted in Fig.~\ref{fig:fig5}.  
As the density grows the orientational order parameter along the aligning direction decreases unlike bulk LC/LCDr system whose global orientational order parameter increases with density\cite{Steuer,lc34}. This happens due to the fact that at lower pressures (lower density) the surface effect is dominant so that almost all the mesogenic units are anchored to the surface and oriented in the aligning direction. In these state points, the system is dictated by the external fields imposed by the walls. As the pressure increases some mesogens and beads leave the surfaces of the walls and form successive layers over the first layer. This phenomenon continues until the space in the box fills. The growth of mesogenic layers along the direction normal to the surface do not guaranty the transmission of mesogenic alignment along the aligning direction. The alignment transmission rate decays with the perpendicular distance from both surfaces of the walls. For instance, the first $2-3$ perpendicular layers are almost perfectly aligned with the field(aligning direction) depending on the temperature. Beyond this distance from both upper and lower walls the alignment of mesogens along the field significantly decreases. 
 
The temperature has adverse effect on global orientational order parameters along each coordinate axes. As the temperature grows, most mesogens, except those in the immediate neighbourhood of the walls, receive adequate thermal energy to get out of surface trap. As the result, their conformation (position and orientation) is governed by the thermodynamic effect enabling them to align randomly in space. A system of randomly oriented mesogenic units ends up with an isotropic value of orientational order parameters.  That is clearly evident from the right panel of Fig.~\ref{fig:fig5}.

\begin{figure}[h!]
\centering
\begin{tabular}{cc}
\includegraphics[width=3.0in,height=1.0in]{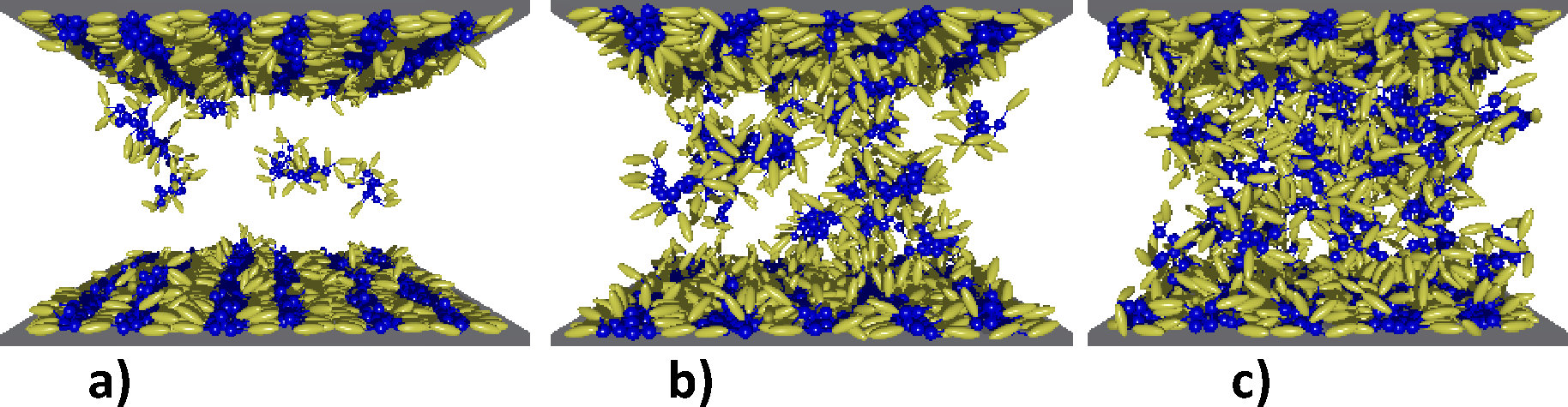}\\
\end{tabular}
\caption{Snapshots of systems of model LCDr confined in a slit-pore under uniform planar anchoring conditions at $P^* = 0.1$. a) at $T^* = 1.0$, b) at $T^* = 1.5$ and c) at $T^* = 2.0$.} 
\label{fig:fig6}
\end{figure}

\begin{figure}[h!]
\centering
\begin{tabular}{cc}
\includegraphics[width=3.0in,height=1.2in]{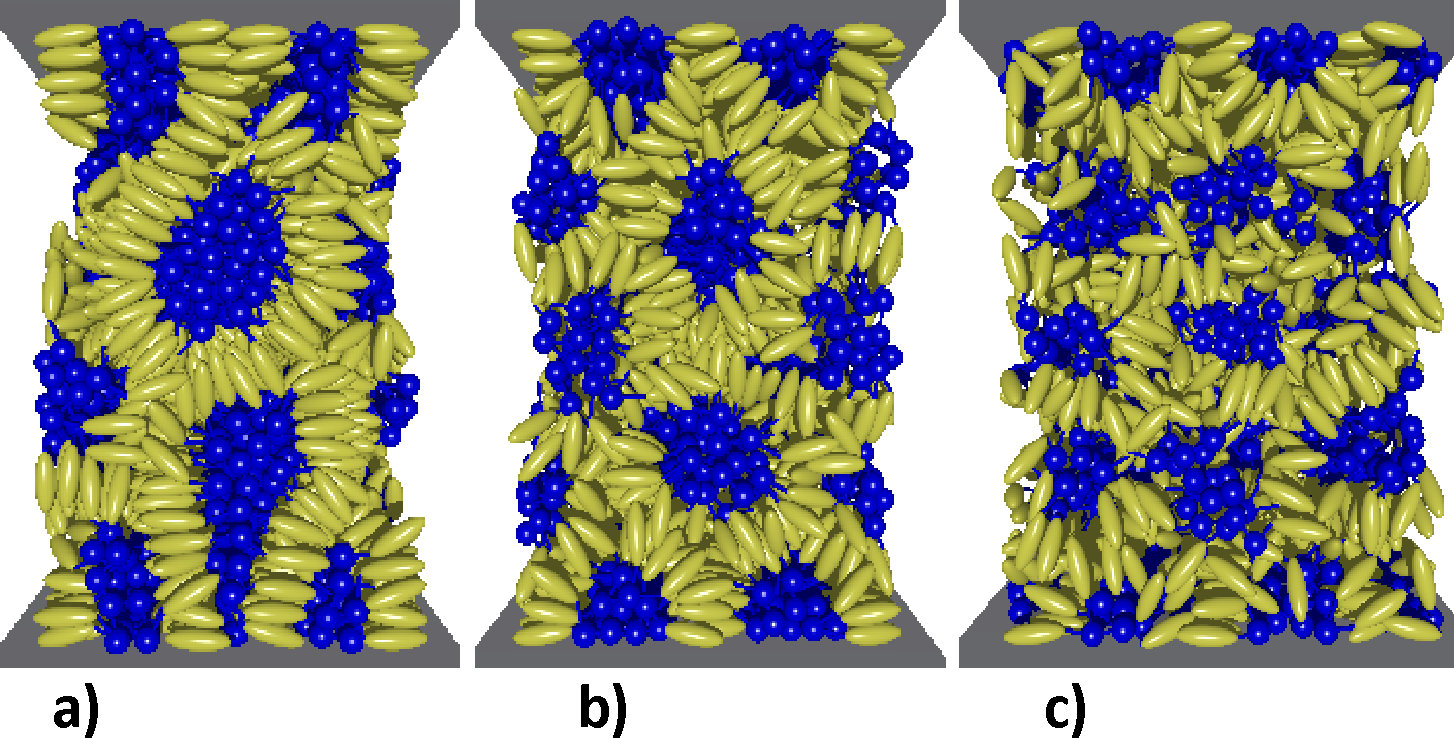}\\
\end{tabular} 
\caption{Snapshots of systems of model LCDr confined in a slit-pore under uniform planar anchoring conditions at $P^* = 1.0$. a) at $T^* = 1.0$, b) at $T^* = 1.5$ and c) at $T^* = 2.0$.}
\label{fig:fig7}
\end{figure}

\begin{figure}[h!]
\centering
\begin{tabular}{cc}
\includegraphics[width=3.0in,height=1.5in]{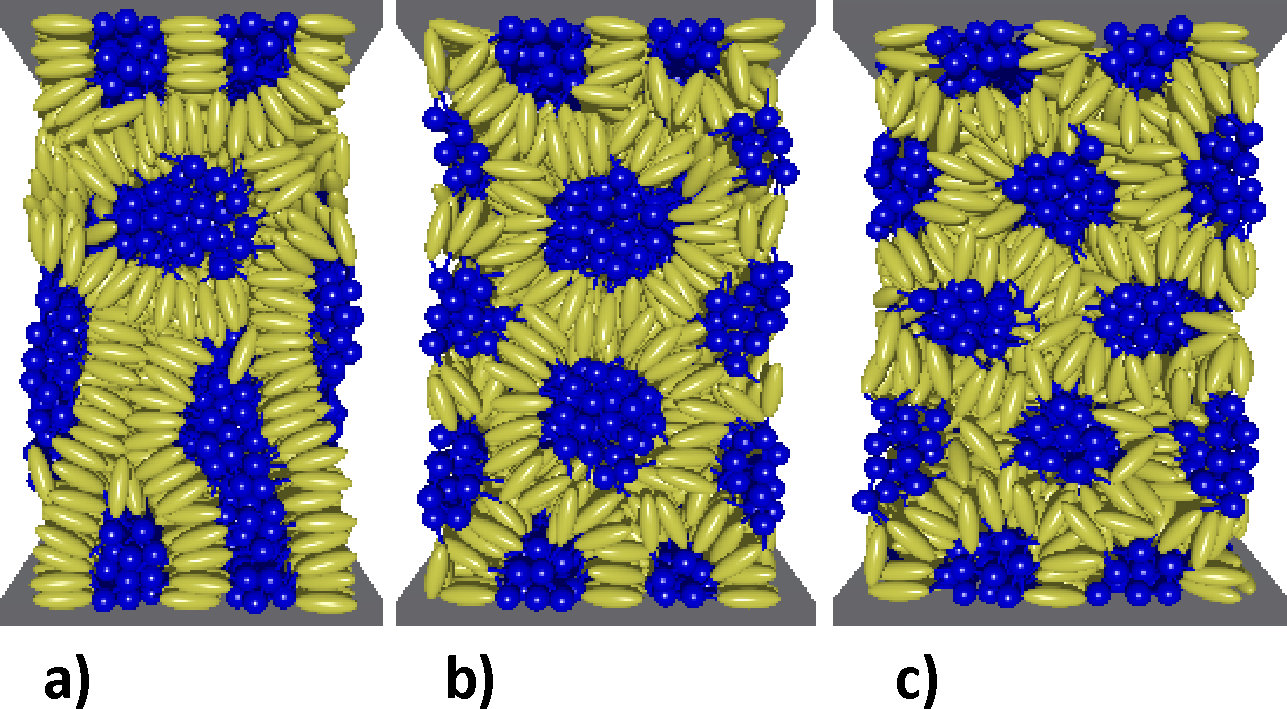}\\
\end{tabular}
\caption{Snapshots of systems of model LCDr confined in a slit-pore under uniform planar anchoring conditions at $P^* = 1.5$. a) at $T^* = 1.0$, b) at $T^* = 1.5$ and c) at $T^* = 2.0$.} 
\label{fig:fig8}
\end{figure}

\begin{figure}[h!]
\centering
\begin{tabular}{cc}
\includegraphics[width=1.2in,height=1.0in]{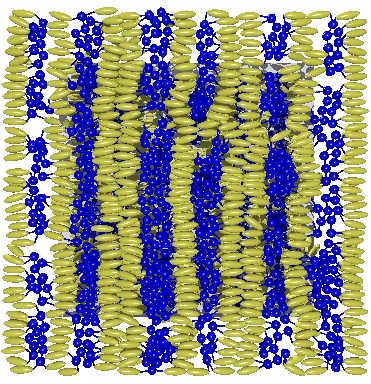}
\includegraphics[width=1.0in,height=1.0in]{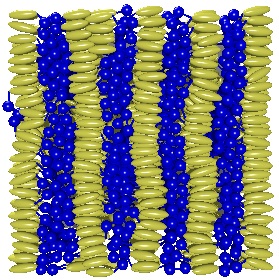}
\includegraphics[width=1.0in,height=1.0in]{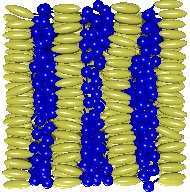}\\
\end{tabular}
\caption{Top or bottom view of a layered structures of LCDrs on the wall (remove the wall for clear vision) in increasing order of pressure $P^*=0.1$, $0.5$ and $1.0$, respectively from left to right at fixed $T^*=1.0$} 
\label{fig:fig9}
\end{figure}

In addition, the outcomes of homogeneous surface-LCDr system simulations are summarized by the snapshots shown in Fig.~\ref{fig:fig6}, ~\ref{fig:fig7}, ~\ref{fig:fig8} and ~\ref{fig:fig9}. The effect of substrate on the configuration of beads and mesogens is readily apparent from all these, with ordered layers of planarly aligned monomers residing nearby the surface of the wall region. Clear separation between beads and mesogens is revealed in the snapshots as expected. This separation could be dictated by either the wall or thermodynamic effect. In the vicinity of the walls, thermodynamic effect is negligible leaving mesogens to align in the aligning direction. However, in the regions far from the walls, typical thermodynamic effect is dominant over the wall effect. In this region, columnar like self-organization of LCDrs are noticed as expected\cite{lc623}. The range of this bulk region depends on both the temperature and pressure. 

At lower pressures a mono-/bi-layers of mesogens and beads along the surface normal and several layers along the aligning direction has been formed. The orientation of most mesogenic units at these pressures is along the aligning direction. At this state, mesogens form well defined $2D$ smectic A like structure with layers of beads in between successive mesogenic layers (see Fig.~\ref{fig:fig9}). As the pressure grows the number of plane-normal layer increases at the expense of parallel layers. Further compression of simulation box results in making some mesogens and beads to leave the previous layers and form the next ones. This phenomena continues until the system is no more compressible. However, the pace of layer transmission is not uniform. As we get far from the surface perpendicular layer formation decays therein creating isotropic state regions around the middle of the slit. 

In order to quantitatively describe the observations revealed in the snapshots, additional averaged quantities have been calculated. These quantities are number density and local orientational order parameters (see Eqs. \ref{eq:eq11} and \ref{eq:eq12})of mesogens along the aligning direction. 
 \begin{figure}[h!]
\centering
\begin{tabular}{cc}
\includegraphics[width=3.2in,height=3.2in]{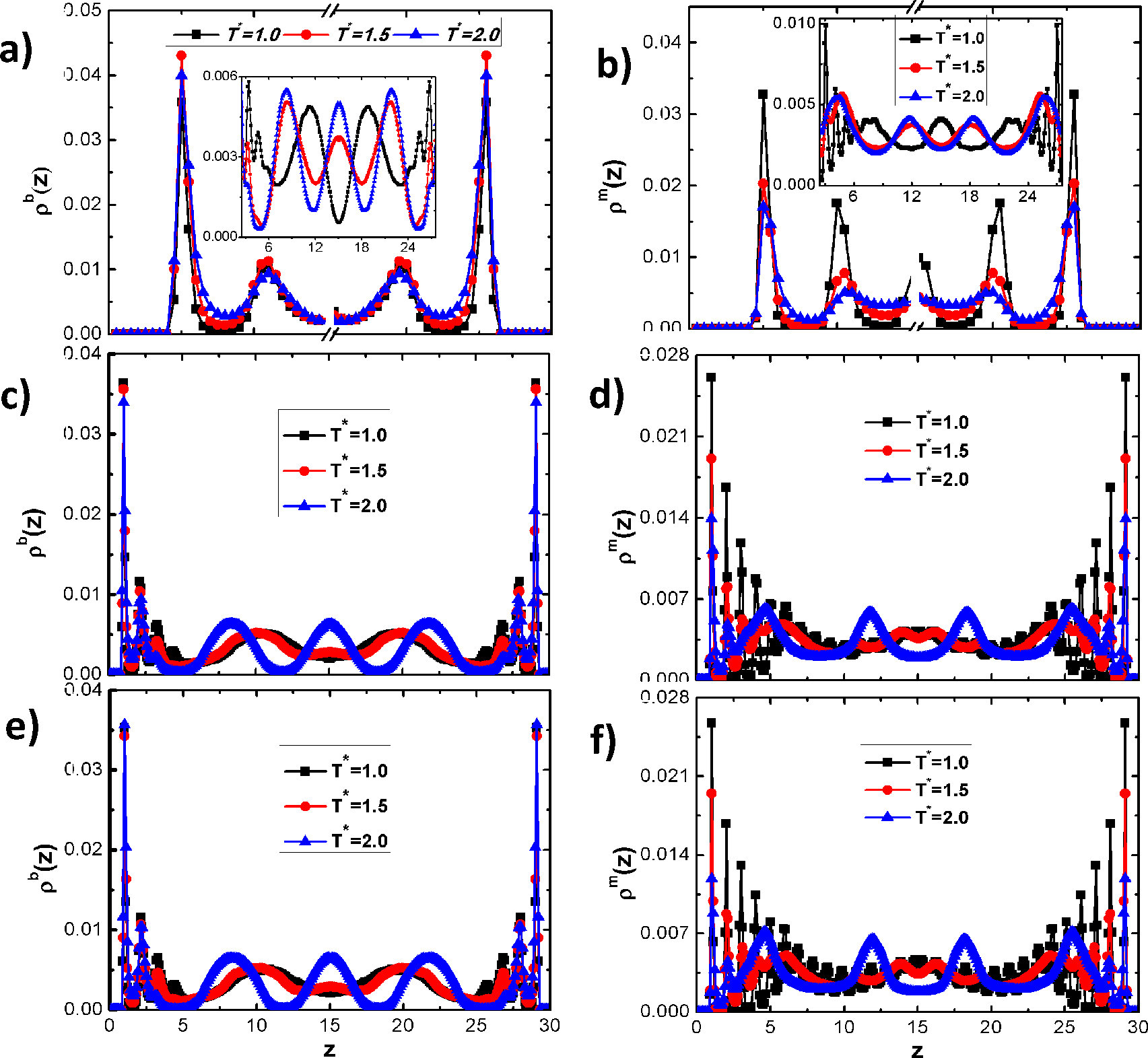}\\
\end{tabular}
\caption{Mass distribution of beads (left column) and mesogenic units (right column) along the surface normal direction, $z-axis$, as the function of distance from the surface at different pressures and temperatures. The rows from top to bottom refer to $P^* = 0.1$, $0.5$ and $1.0$, respectively.}
\label{fig:fig10} 
 \end{figure} 
The density profiles depicted in Fig.~\ref{fig:fig10} show the
adsorption characteristics for the film confined
between the homogeneous planar surface regions as the functions of distance along  surface normal direction. It is immediately apparent from this figure that layer propagation is highly dumped when the distance from the surface increases, specially, in lower pressure cases. In addition, different peak values at the surfaces of the walls at different pressures tells us that some of the monomers have left the surface as the result of increase in pressure. This justifies what has been seen in Fig.~\ref{fig:fig9} where clear decrease in in-plane number of layers is realized.  Increasing pressure leads to formation of layers one over the other with a periodicity of $\sigma_{0bb}$ (for beads) and $\sigma_{0mm}$ (for mesogens). Surface induced mesogenic layers can traverse the whole region of the slit at higher pressures with dumped like oscillation going from the surface to the middle region. However, there is no significant change in the distribution of beads as the function of distance from the surface with the pressure.
  
\begin{figure}[h!]
\centering
\includegraphics[width=3.0in,height=1.5in]{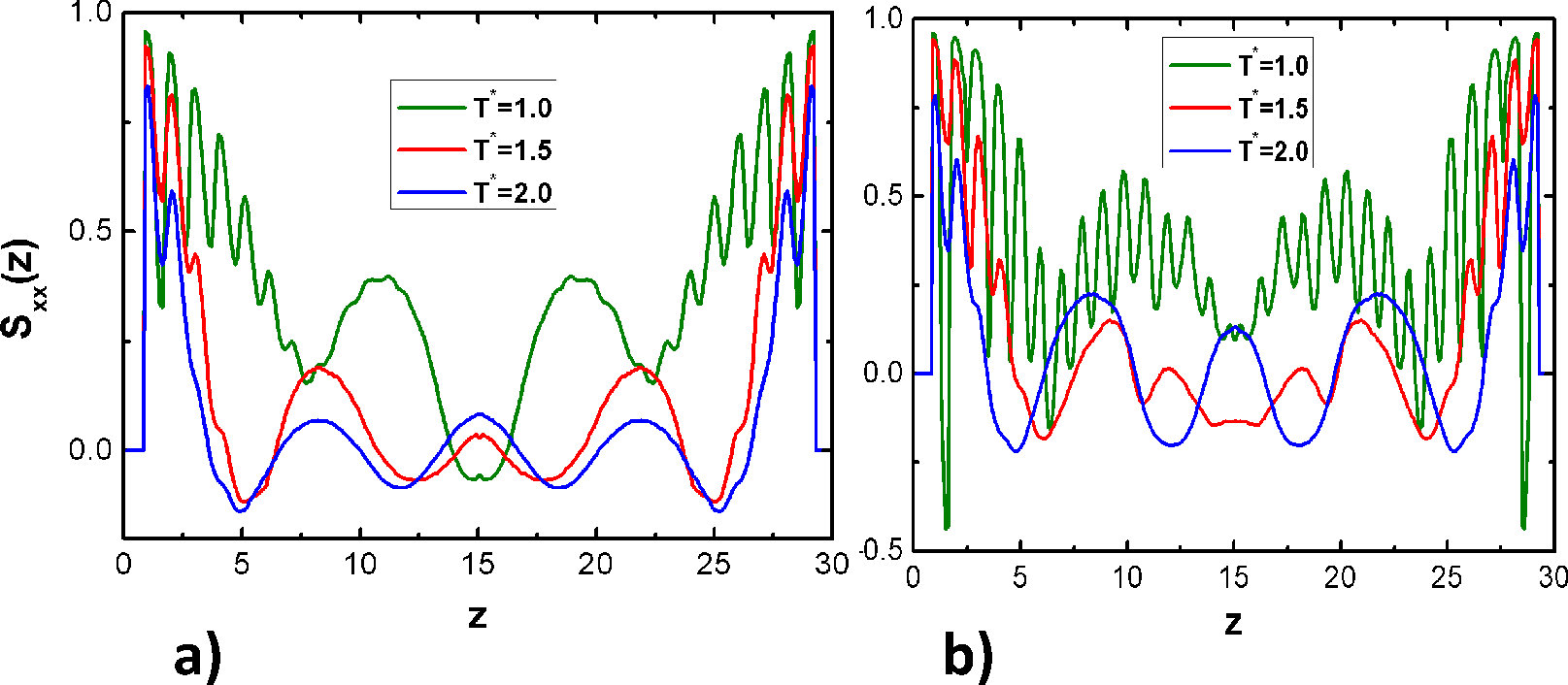}\\
\caption{Average local orientational order parameter along the aligning direction as the function of perpendicular distance $z$ from the surface at $P^*=0.5$ (a) and $P^* = 1.0$ (b).} 
\label{fig:fig11}
 \end{figure}

 A more complete understanding of the orientational aspects of the substrate-induced ordering in this system can be obtained from the local components of orientational order tensor; $S_{xx}(z)$ or $S_{zz}(z)$ (diagonal components of the order tensor). For perfect homogeneous planar alignment, $S_{xx}(z)$ should tend to $1$ and $S_{zz}(z)$ should tend to $-0.5$. Fig.~\ref{fig:fig11} shows $S_{xx}(z)$ profiles as the function of perpendicular distance from the surfaces of the walls. As predicted, the value of $S_{xx}(z)$ is nearly $1$ in the first $2-3$ layers from both top and bottom walls for all pressure and temperature values. After these layers, it decreases even up to negative value highly depending on the pressure and temperature. At lower temperature and higher pressure there looks a better alignment transmission from the surface to the bulk but still not uniform. Even much lower temperature($T^*=0.5$) simulation has been performed to test if there is any better alignment transmission but we end up with no significant difference from that of $T^*=1.0$. Consequently, achieving a uniform perpendicular layer of mesogens of LCDrs oriented in the aligning direction looks challenging task at any surface-LCDr coupling or thermodynamic conditions. 

\section{\label{sec:level1}Conclusions}
A coarse-grained computational simulation study of $G_2D_3$ model LCDr system confined in a slit-pore of two parallel unpenetrable walls  under uniform planar anchoring condition has been carried out. The simulation results show significant conformational and structural changes driven by the effect of the substrate at each state points. Substrate induced smectic-A like layers of mesogenic units separated by layers of dendritic core (beads) has been observed along the aligning direction. In the direction normal to the plane of the surface, the development of mesogen layer formation is pressure dependent. At lower pressure, there was only single layers of mesogens and beads in the vicinity of the substrates. As the pressure grows, number of perpendicular mesogenic layers grows. This transmission of layer formation or alignment of mesogens in the direction perpendicular to the surface is not uniform, but, it decays as it goes from the vicinity to the middle of the slit. In the regions where smectic-A like structure of mesogenic units are apparent, mesogens from a single LCDr contribute to smectic layers in the left and right side of dendrimer core (beads). According to simulation results strong coupling between internal molecular structure and molecular environment is identified. This leads to mesogenic groups becoming aggregated at the two ends of LCDrs in the regions where the effect of the walls is dominant. Consequently, individual LCDr molecules change their shapes from "spherical" in to "rod" shape.

The LCDr studied here could represent a wider range of polyphilic mesomorphic materials in which separate building blocks with different types of interaction sites are combined. Substrate induced micro-phase separation and alignment transmission in these materials opens up the opportunity to self-assemble in to well-defined nanostructures for different potential applications. Potential applications could be molecular electronics, opto-electronics, photonics, nanotechnology, etc. The CG-simulation method presented here can be used as an inspiration towards further simulation and experimental studies on these systems. As a result, it could be a useful tool in helping design confined nanoscale structures for future applications.   

\bibliographystyle{rsc}
\bibliography{reference}

\end{document}